\documentclass[graybox]{svmult}

\usepackage{helvet}         % selects Helvetica as sans-serif font
\usepackage{courier}        % selects Courier as typewriter font
\usepackage{type1cm}        % activate if the above 3 fonts are not available on your system

\usepackage{makeidx}         % allows index generation
\usepackage{graphicx}        % standard LaTeX graphics tool when including figure files
\usepackage{multicol}        % used for the two-column index
\usepackage[bottom]{footmisc}% places footnotes at page bottom

\usepackage{color}
\definecolor{keywordcolor}{rgb}{0,0,1}

\begin{document}

%\title*{Single photon absorption by a single trapped atom: a key requirement for quantum networking}
\title*{Single photon absorption by a single atom: from heralded absorption to polarization state mapping}
\titlerunning{Single photon absorption by a single atom}
\author{Nicolas Piro and J\"urgen Eschner}
\institute{Nicolas Piro \at \'Ecole Polytechnique F\'ed\'erale de Lausanne, \email{nicolas.piro@gmail.com}
\and J\"urgen Eschner \at Universit\"at des Saarlandes, \email{juergen.eschner@physik.uni-saarland.de}}
\maketitle

\abstract{Together with photon emission, the absorption of a single photon by a single atom is a fundamental process in matter-light interaction that manifests its quantum mechanical nature. As an experimentally controlled process, it is a key tool for the realization of quantum technologies. In particular, in an atom/photon based quantum network scenario, in which localized atomic particles are used as quantum information processing nodes while photons are used as carriers of quantum information between distant nodes, controlling both emission and absorption of single photons by single atoms is required for quantum coherent state mapping between the two entities. Most experimental efforts to date have focused on establishing the control of single photon emission by single trapped atoms, and the implementation of quantum networking protocols using this interaction. In this chapter, we describe experimental efforts to control the process of single photon absorption by single trapped ions. We describe a series of experiments in which polarization entangled photon pairs, generated by a spontaneous parametric down-conversion source, are coupled to a single ion. First the source is operated to generate heralded single photons, and coincidences between the absorption event of one photon of the pair and the detection of the heralding partner photon are observed. We then show how polarization control in the process is established, leading to the manifestation of the photonic polarization entanglement in the absorption process. Finally, we introduce protocols in which this interaction scheme is harnessed to perform tasks in a quantum network, such as entanglement distribution among distant nodes of the network, and we demonstrate a specific protocol for heralded, high-fidelity photon-to-atom quantum state transfer.}

\section{Introduction} \label{sec:Introduction}

At its most fundamental level, matter-light interaction involves the absorption and emission of single photons by single atomic particles. While being well understood theoretically, these processes have only recently started to be explored experimentally under controlled conditions. The motivation for realizing photon-atom interaction experiments at the single particle level is two-fold: first, from the fundamental perspective, such studies allow performing tests of the quantum theory describing the interaction; second, controlling the interaction opens new routes to developing technologies that use quantum mechanical phenomena as a resource, in particular for purposes of quantum information processing.

Experiments with single trapped atomic ions have proven to provide optimal conditions for quantum information processing, fulfilling the requirements of providing long-coherence qubits \cite{Langer2005}, high-fidelity state manipulation and detection schemes \cite{LeibriedReview2003}, as well as controlled coherent interaction of several quantum bits allowing deterministic generation of entanglement \cite{Turchette1998}, quantum logic gates \cite{Monroe1995} and quantum error correction \cite{Chiaverini2004}. Meanwhile, single photons are optimal carriers of quantum information, allowing the transfer of quantum states and the distribution of entanglement over long distances \cite{Ursin2007}. The quantum-coherent link between single atomic particles and single photons then emerges as a key requirement to develop quantum networks, in which quantum information is processed and stored in atom-based quantum nodes, while photons are used to communicate quantum states between the nodes \cite{Cirac1997, Kimble2008}. 

Several experiments have concentrated on establishing quantum-coherent links between single trapped atoms and emitted photons, taking advantage of the intrinsic quantum correlations in the emission process \cite{Blinov2004, Volz2006}. Recent experiments even showed how to entangle distant trapped atomic particles by allowing the emitted photons to interfere and subsequently performing a correlation measurement on the photons \cite{Moehring2007, Hofmann2012}. However, a fully bidirectional link requires also control over the absorption process in a quantum coherent fashion. Here we concentrate on recent progress with single ions and single photons. A cavity-based approach with neutral atoms is also being pursued \cite{Ritter2012}. A bidirectional link has also been established between atomic ensembles \cite{Chaneliere2005}, but these do not possess the quantum computation capabilities provided by ion chains.

The control of single photon absorption not only allows one to transfer quantum states encoded in a photonic degree of freedom onto an internal degree of freedom of the atom. It also enables taking advantage of the widely developed quantum photonic technologies, such as quantum light sources, in the realm of atom-based quantum networks. In particular, one can think of implementing protocols to entangle distant atoms by generating the entanglement via optical means, e.g. using the process of spontaneous parameteric down-conversion (SPDC), and subsequently mapping the states of the two photons onto two distant atoms by an entanglement swapping process. Such absorption-based schemes could benefit from the high generation rates for entangled photon pairs available in state-of-the-art sources and thereby provide an attractive alternative to emission-based schemes.

In this context, we review in this chapter a series of experiments in which entangled photon pairs, generated with an SPDC source, interact with a single trapped $^{40}$Ca$^+$ ion. The spectral properties of the photons are tailored to optimally interact with the single atom, thereby enabling several photon-atom entanglement schemes. 

We first present in Section \ref{sec:InteractionSchemes} schemes that allow one to observe single atom-photon interaction processes, to map quantum states from photons to $^{40}$Ca$^+$ ions and to perform entanglement swapping between photon pairs and distant trapped ions.  In Section \ref{sec:Setup} we describe the experimental apparatus we developed to implement and study these interaction schemes, consisting of a pair of $^{40}$Ca$^+$ ion trap setups and an SPDC photon pair source designed to produce narrowband polarization-entangled photon pairs tunable to two optical transitions in $^{40}$Ca$^+$.  We present the results of various interaction experiments in Section \ref{sec:Results}.  Finally, in Section \ref{sec:ConclusionAndOutlook} we discuss future possibilities of these experiments and the conclusions of this research. In an appendix 
%\ref{sec:DataAnalysis}
we provide some details regarding the data analysis techniques used to detect the photon-atom interaction.

\section{Single photon - single atom interaction and entanglement schemes}\label{sec:InteractionSchemes}

\begin{figure}[t!]
\centering
  \includegraphics[width=0.8\textwidth]{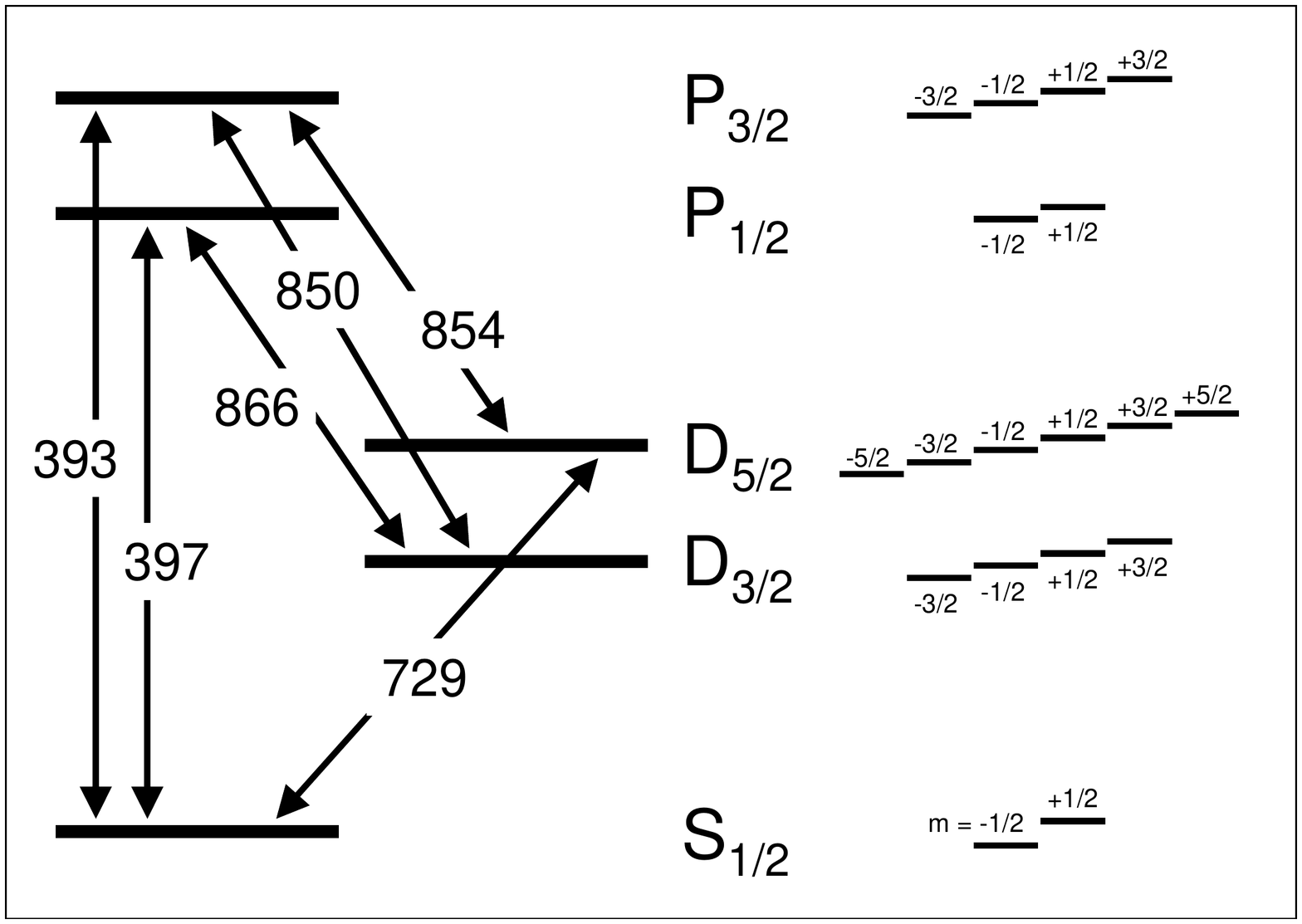}
  \caption{Level scheme and transition wavelengths (given in nm) of the $^{40}$Ca$^+$ ion, without (left) and with (right) magnetic sublevels. Splittings are not to scale; the $g_J$-factors of the states, from bottom to top, are $(2,\frac{4}{5},\frac{6}{5},\frac{2}{3},\frac{4}{3})$. The D levels are metastable with $\sim1$~s natural lifetime. The excited levels  P$_{1/2}$ and P$_{3/2}$ decay preferentially to the ground state manifold S$_{1/2}$, emitting a 397 and 393~nm photon, respectively. The lower excited state P$_{1/2}$ also decays to D$_{3/2}$, with about 6.4\% branching ratio. The upper excited state P$_{3/2}$ also decays to D$_{5/2}$ or D$_{3/2}$, with branching ratios of about 5.9\% and 0.66\%, respectively. Magnetic sublevels of S$_{1/2}$ and D$_{5/2}$ are used to form optical qubits, which are coherently manipulated by laser excitation of the 729~nm electric quadrupole transition. The sublevels of S$_{1/2}$ form a Zeeman qubit manipulated by radio frequency (RF) radiation.\label{fig:LevelScheme}}
\end{figure}

The detailed level scheme, including the Zeeman structure, of $^{40}$Ca$^+$ is depicted in Figure \ref{fig:LevelScheme}. For quantum information processing applications, two types of qubits can be chosen: an optical qubit formed by one magnetic sublevel of S$_{1/2}$ and one sublevel of D$_{5/2}$, or a Zeeman (or RF-) qubit formed by the two sublevels of S$_{1/2}$. The entanglement transfer schemes considered in this work use the latter.

\subsection{Single photon absorption schemes}

\begin{figure}[t!]
\centering
  \includegraphics[width=1\textwidth]{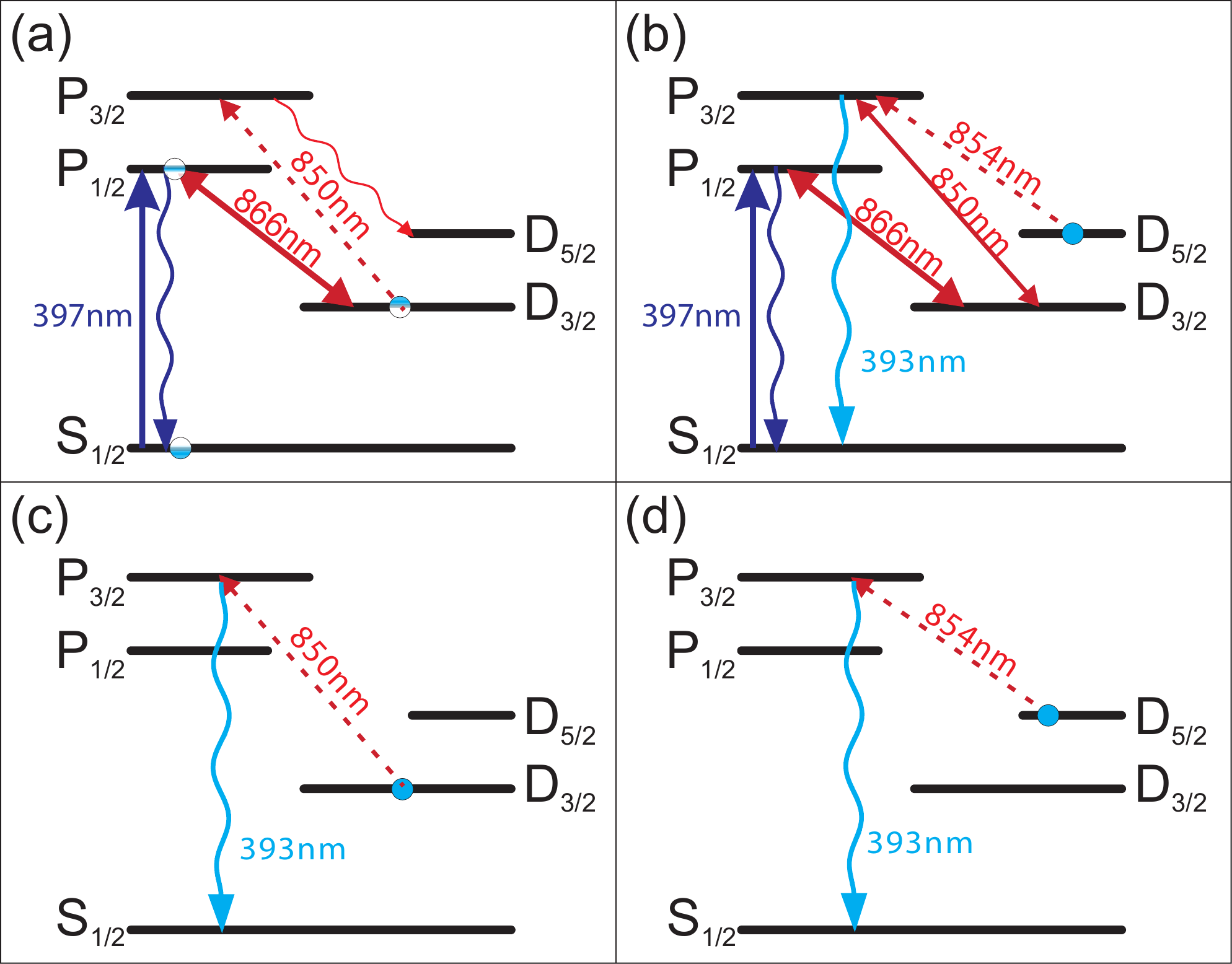}
  \caption{Schemes to detect the absorption of a single photon by a single $^{40}$Ca$^{+}$ ion. Absorption happens either on the 854~nm (b,d) or the 850~nm transition (a,c); it is signaled by either a quantum jump (a,b) or a single emitted photon (c,d). See main text for more details.\label{fig:InteractionSchemes}}
\end{figure}

We consider here four possible schemes addressing either the D$_{3/2} \leftrightarrow$ P$_{3/2}$ transition at 850~nm wavelength or the D$_{5/2} \leftrightarrow$ P$_{3/2}$ transition at 854~nm, see Figure \ref{fig:InteractionSchemes}. 

Schemes (a) and (b) are based on the detection of quantum jumps: the ion undergoes continuous excitation by a 397~nm laser driving the S$_{1/2} \leftrightarrow$ P$_{1/2}$ transition (and at the same time Doppler-cooling the ion) and an 866~nm laser driving the D$_{3/2} \leftrightarrow$ P$_{1/2}$ transition and repumping the ion out of the metastable D$_{3/2}$ level. In this closed cycle, the ion emits a continuous stream of 397~nm fluorescence photons, typically at a few $10^5 ~{\textrm s}^{-1}$ rate, that are recorded with a single-photon counter or a CCD camera. In scheme (a), the single photon source is tuned to the 850~nm line, and the absorption of a photon, followed by spontaneous decay into the metastable D$_{5/2}$ state, puts the ion out of the 397/866~nm transition cycle, hence halting the emission of 397~nm fluorescence. This sudden drop in fluorescence is termed a bright-to-dark quantum jump. In this case it signals the absorption of a single 850~nm photon by the atom. In scheme (b), the photon source is tuned to the transition at 854~nm. An additional weak laser at 850~nm is introduced to induce transitions of the ion into the D$_{5/2}$ state, in which fluorescence emission is suppressed. Now, the absorption of a single 854~nm photon followed by a spontaneous decay to S$_{1/2}$ or D$_{3/2}$ reinitiates the stream of fluorescence emission. Hence the absorption of a single photon is signaled by the onset of fluorescence, or a dark-to-bright quantum jump.

In schemes (c) and (d), the ion is previously prepared in one of the metastable states D$_{3/2}$ or D$_{5/2}$ by optical pumping techniques or by coherent excitation, as described in section \ref{sec:Results}. During the exposure to the single photon source, no other lasers drive the ion. The absorption of an 850 or 854~nm photon, in each case, is now signaled by the emission of a single 393 nm photon, when this is detected by a single photon counter. As we will see, schemes (c) and (d) have the advantage that they do not destroy the final state of the ion and thus enable quantum states to be transferred from the photon to the ion.

While schemes for detecting 850 and 854~nm photon absorption events are both feasible, the 854~nm schemes are significantly more efficient due to a combination of two factors: firstly, the oscillator strength of the D$_{5/2} \leftrightarrow $ P$_{3/2}$ transition is about 6 times larger than that of the D$_{3/2} \leftrightarrow $ P$_{3/2}$ transition; secondly, upon excitation to P$_{3/2}$, the probability of decaying to D$_{5/2}$ (required by scheme (a)) is only 5.9\% \cite{Gerritsma2008}, while decay to S$_{1/2}$ (required in (b)) happens with nearly 94\% probability. Overall, scheme (b) is about 100 times more efficient the scheme (a). The first factor also makes scheme (d) about one order of magnitude more efficient than scheme (c).

\subsection{Photon-atom state transfer and entanglement swapping schemes} \label{subsec:PhotonAtomInteraction}

Entanglement swapping schemes between photons and atoms rely on the possibility of mapping the polarization state of the photon onto the internal state of the ion. In particular, an arbitrary photon state $\left|\Psi_{p}\right\rangle = \alpha\left|\uparrow\right\rangle + \beta\left|\downarrow\right\rangle$ must be transferred to the state $\left|\Psi_{i}\right\rangle = \alpha\left| + \right\rangle+\beta\left|-\right\rangle$, where $\left|\uparrow\right\rangle$, $\left|\downarrow\right\rangle$ are two arbitrary polarization basis vectors, and $\left|+\right\rangle$, $\left|-\right\rangle$ represent an atomic qubit pair of states.

\begin{figure}[t!]
\centering
  \includegraphics[width=1\textwidth]{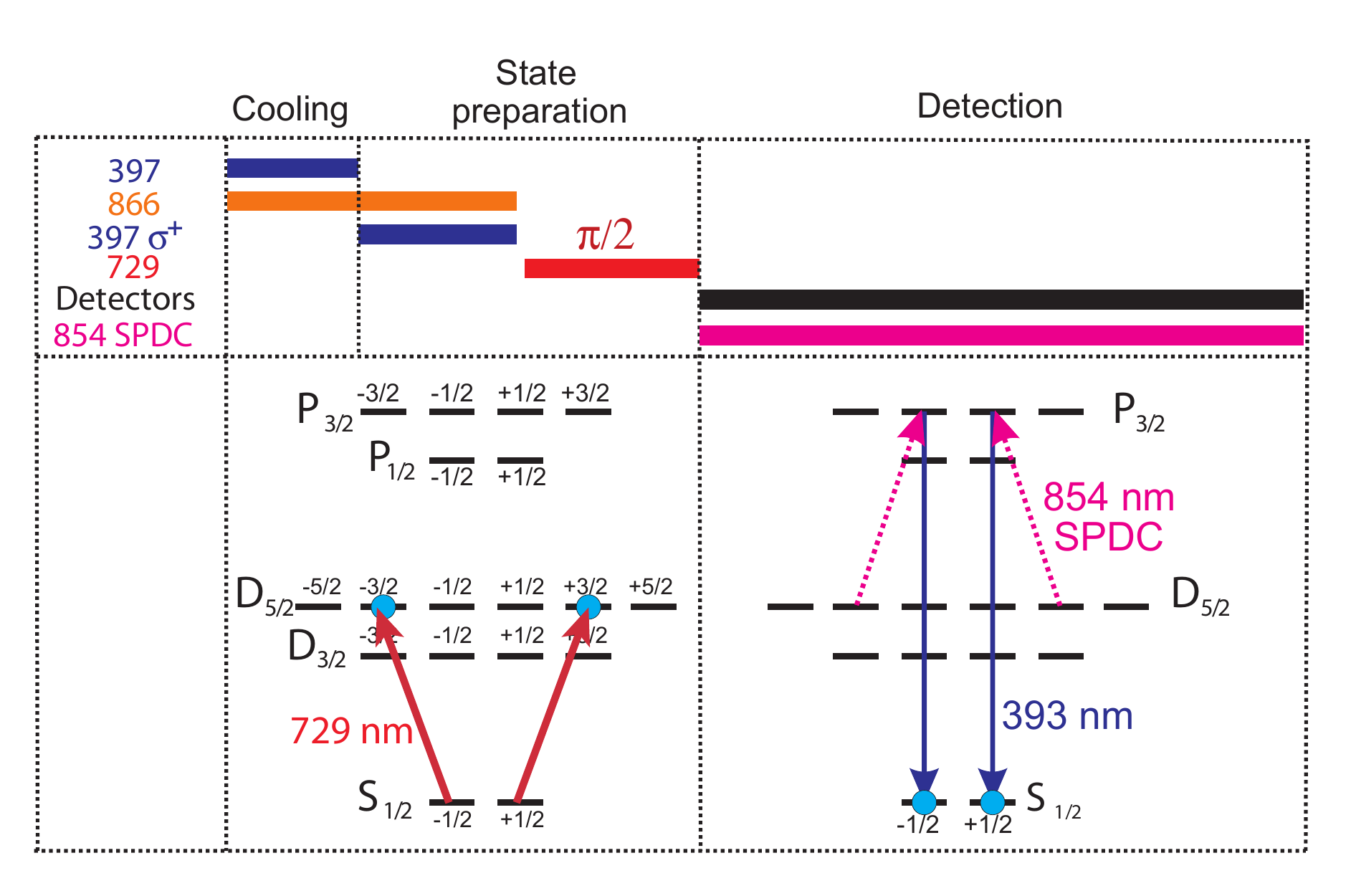}
  \caption{Scheme to transfer the polarization state of an 854 nm photon onto the atom  \cite{Mueller2014, Kurz2014}.\label{fig:StateTransferScheme854}}
\end{figure}

One possible proposed state transfer scheme is illustrated in Figure \ref{fig:StateTransferScheme854} \cite{Mueller2014, Kurz2014}. Here the atomic qubit is chosen to be $\left|\pm\right\rangle = \left|{\textrm S}_{1/2}, m = \pm 1/2\right\rangle$. The ion is manipulated with a laser pulse sequence consisting of three main phases. It is first laser cooled by means of the 397 and 866~nm lasers as usual. Then it is prepared in a symmetric superposition of the states $\left|{\textrm D}_{5/2}, m=\pm 3/2\right\rangle$ by means of four pulses: first, a $\sigma^{-}$ polarized 397~nm beam pumps the ion to the $\left|{\textrm S}_{1/2}, m= -1/2\right\rangle$ state; then, a resonant RF $\pi/2$-pulse creates a symmetric superposition of the two sublevels of S$_{1/2}$; finally, two 729~nm $\pi$-pulses coupling the respective S$_{1/2} \leftrightarrow$ D$_{5/2}$ transitions convert this state into the desired $\left|{\textrm D}_{5/2}, m=\pm 3/2\right\rangle$ superposition. 

%The pulse must contain a coherent superposition of the two frequencies coupling $m_S = -1/2 \leftrightarrow m_D = 3/2$ and $m_S = -1/2 \leftrightarrow m_D = -3/2$ !!!DOES NOT CORRESPOND TO FIGURE!!!. Note that for this $|\Delta m| = 2$ transitions are required, which is possible thanks to the quadrupole character of the S-D transition.

At this point, the ion is exposed to a single photon at 854~nm, for example from an SPDC source, and a single photon detector monitoring the 393~nm photons is gated on. The SPDC photon is prepared in an arbitrary polarization state $\left|\Psi_{p}\right\rangle = \alpha\left|H\right\rangle+\beta\left|V\right\rangle$; upon its absorption, the ion undergoes a Raman transition to the ground state emitting a 393~nm photon which, if detected, heralds the absorption event. The process is coherent, i.e., the absorbed polarization is converted into a corresponding superposition of the RF-qubit in S$_{1/2}$, as long as it remains impossible to determine which P$_{3/2}$ Zeeman substate the ion decays from. This condition holds if the 393~nm photon is detected with a time resolution whose inverse is much larger than the frequency splitting between the two involved, equally polarized P$_{3/2} \to {\textrm S}_{1/2}$ transitions \cite{Togan2010}. Such high time resolution is also needed to keep track of the phase of the final superposition state in S$_{1/2}$, see \cite{Kurz2014, Schug2014}.

\section{Experimental setup}\label{sec:Setup}

\begin{figure}[t!]
\centering
  \includegraphics[width=0.9\textwidth]{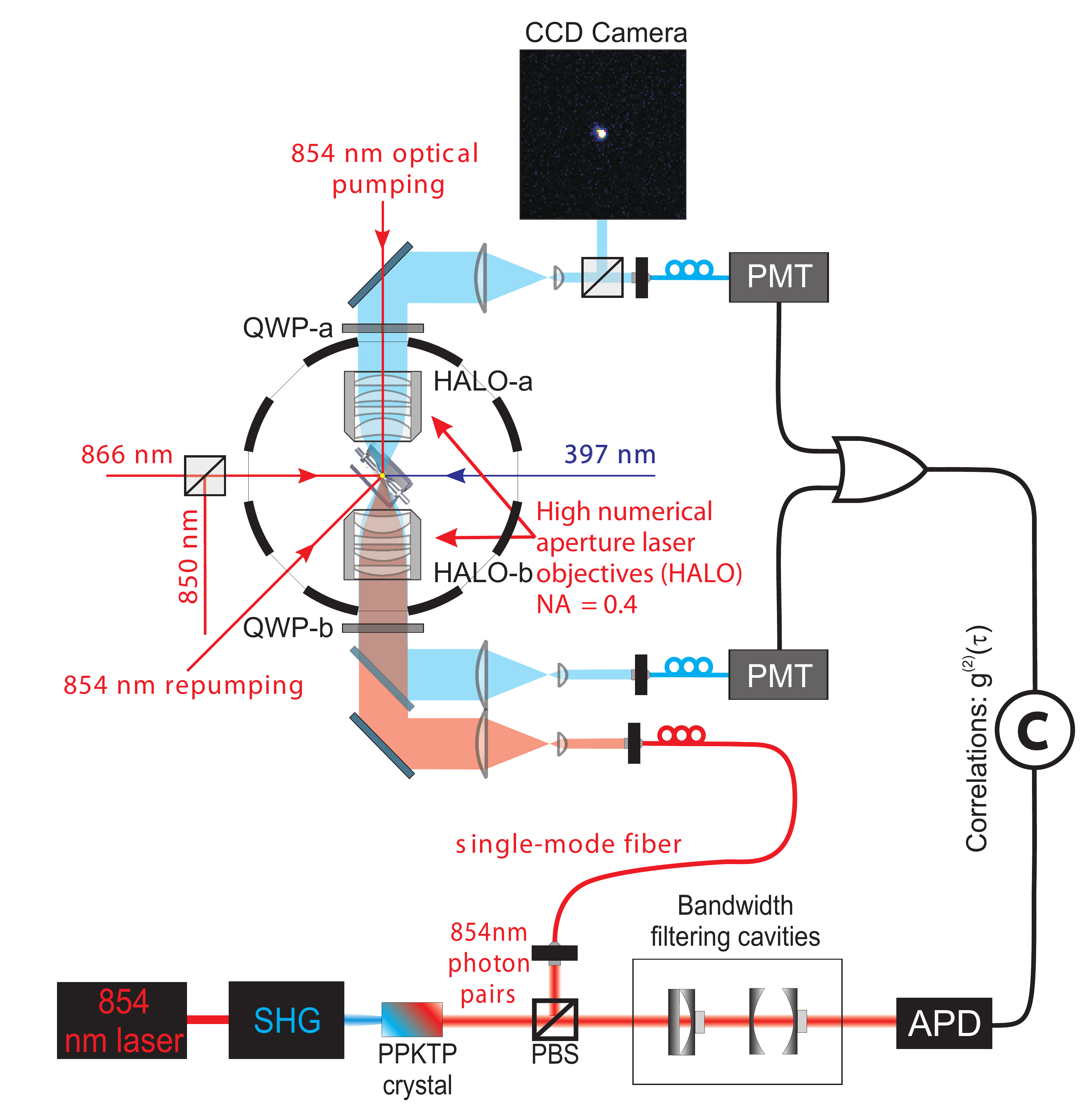}
  \caption{Experimental setup combining an ion trap with an SPDC photon source, the latter being operated as a heralded single-photon source. Two high-numerical aperture laser objectives (HALOs) facilitate the ion-photon coupling. See text for more details. \label{fig:IonSourceSetup}}
\end{figure}

The core of the experimental setup is a twin ion trap apparatus capable of trapping single or strings of  $^{40}$Ca$^+$ ions. For details on their construction and operation, see Refs. \cite{Rohde2009Thesis, Schuck2010Thesis, Almendros2009}. These setups are complemented with a source of polarization entangled photon pairs tunable to two transitions in $^{40}$Ca$^+$, described in Refs. \cite{Haase2009, Piro2009}. Here we focus on the operation of one ion trap setup in conjunction with the single photon source, see Figure \ref{fig:IonSourceSetup}. 

The SPDC photon pair source is based on a laser at 854~nm whose frequency-doubled light pumps a 20-mm long PPKTP crystal designed for degenerate type-II spontaneous parametric down-conversion. As the key purpose of the source is to produce photon pairs with spectral properties matching those of atomic transitions in the $^{40}$Ca$^+$ species, the wavelength of the 854 nm laser is actively locked to the transition, and the photon bandwidth is matched to the atomic absorption linewidth of 22~MHz by means of a cavity filtering system. Photon pairs are generated withing a relatively large bandwidth of about 200 GHz. However, due to the narrow linewidth of the pump laser and energy conservation in the SPDC process, the pairs are emitted at symmetric frequencies with respect to the pump, $\omega_s = \omega_p - \omega_i$. Hence, filtering one photon of the pair with filtering cavities locked on resonance to the 854 pump laser, and a design transmission bandwidth of 22 MHz, followed by the detection of the transmitted photon heralds the presence of a partner photon resonant with the atomic transition \cite{Haase2009, Piro2009}. The partner photon is coupled into a single mode fiber and directed to the ion trap setup. The photon pair exhibits high-purity polarization entanglement which may be utilised when the polarizing beam splitter separating the pair (PBS in Fig.~\ref{fig:IonSourceSetup}) is replaced by a non-polarizing beam splitter \cite{Haase2009, Piro2009}.

The ion trap setup consists of a linear Paul trap placed between two high-numerical aperature lenses (HALO-a/b in Fig.~\ref{fig:IonSourceSetup}) designed to be diffraction limited and near-achromatic at wavelengths around 397 and 860 nm, with 0.4 numerical aperture. The HALO lenses serves two purposes: to efficiently collect light scattered by the ion, and to tightly focus the single photon beam onto the ion to enhance its absorption probability. Light from the source is first expanded by a magnifying telescope to fill the back-aperature of the objective, and aligned to focus at the ion position. Fluorescence light from the ion is collected by both objectives and detected by two independent photo-multiplier tubes (PMT), or by a CCD camera (on one side) to image the ion. Photon detection pulses from both PMTs are combined into a single channel. Their detection times, together with the detection times of the source herald photon, are recorded in a computer by means a time-tagged single-photon counting system (PicoHarp).

\section{Experimental progress}\label{sec:Results}

In this section we present the experimental progress towards establishing a quantum coherent link between single photons and a single atom. Specifically, we first describe a preliminary experiment in which scheme (b) of Fig.~\ref{fig:InteractionSchemes} is implemented in order to detect the absorption of weak 854~nm light by a single $^{40}$Ca$^+$ ion. We then present an experiment where we detect time correlations between the absorption event and the heralding partner photon, establishing the true single-photon nature of the absorption process. In a further experiment, we show how control of the Zeeman substate of the atoms renders the absorption dependent on the polarization of the photon, hence manifesting the polarization entanglement of the photon pair in the absorption process. Finally, we discuss the experimental implementation of a heralded-absorption protocol through which the polarization state of a single absorbed photon is transferred, with high fidelity, onto the qubit state of the ion.

\subsection{Single photon absorption by a single ion}

As described in section \ref{sec:InteractionSchemes}, the absorption of a single photon by a single atom may be signaled by a change in its fluorescence state, a quantum jump \cite{Nagourney1986, Sauter1986, Bergquist1986}. This uses the concept of quantum amplification proposed by Dehmelt as early as 1975 \cite{Dehmelt1975}. In the case of the scheme in Fig.~\ref{fig:InteractionSchemes}(b), the atom starts emitting fluorescence photons upon the absorption of a single 854~nm photon, providing a very efficient signal that witnesses the absorption event.

\begin{figure}[t!]
\centering
  \includegraphics[width=\textwidth]{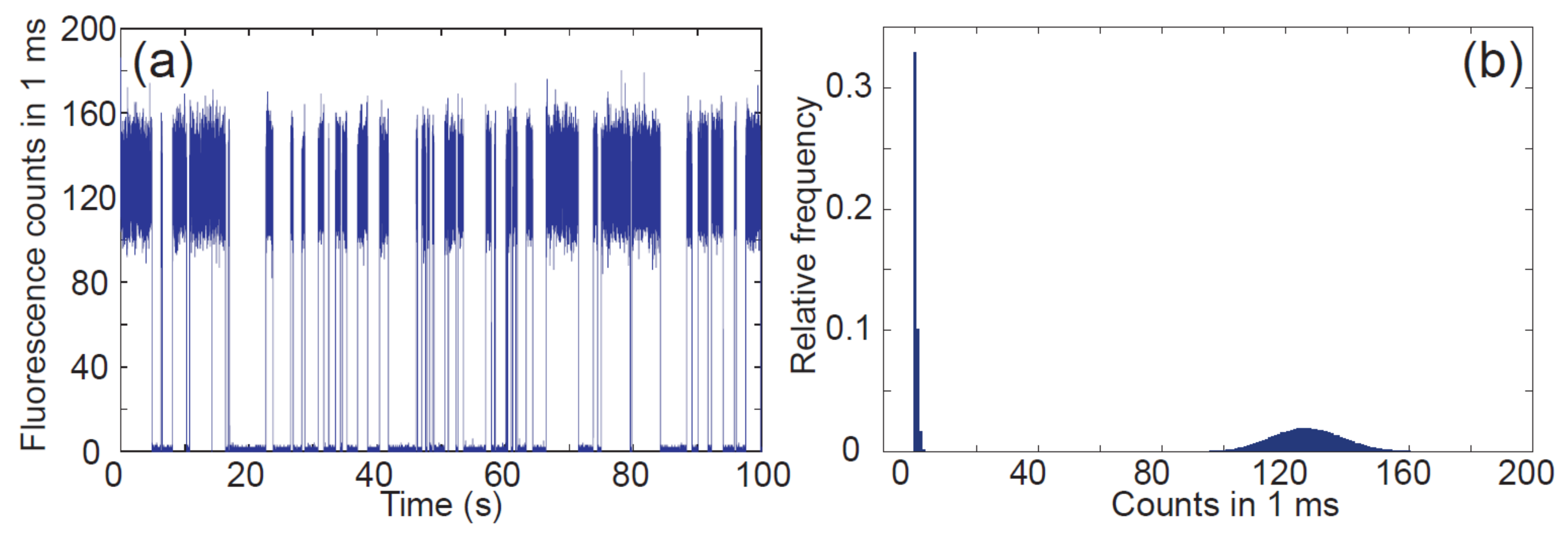}
  \caption{(a) Detected fluorescence rate (in time bins of 1~ms) displaying quantum jumps upon excitation with the cooling (397 nm) and repumping (866 nm) lasers, and a weak optical pumping laser (850 nm). (b) Histogram of fluorescence counts displaying the distributions for the on and off states.\label{fig:QJdata}}
\end{figure}

\begin{figure}[t!]
\centering
  \includegraphics[width=\textwidth]{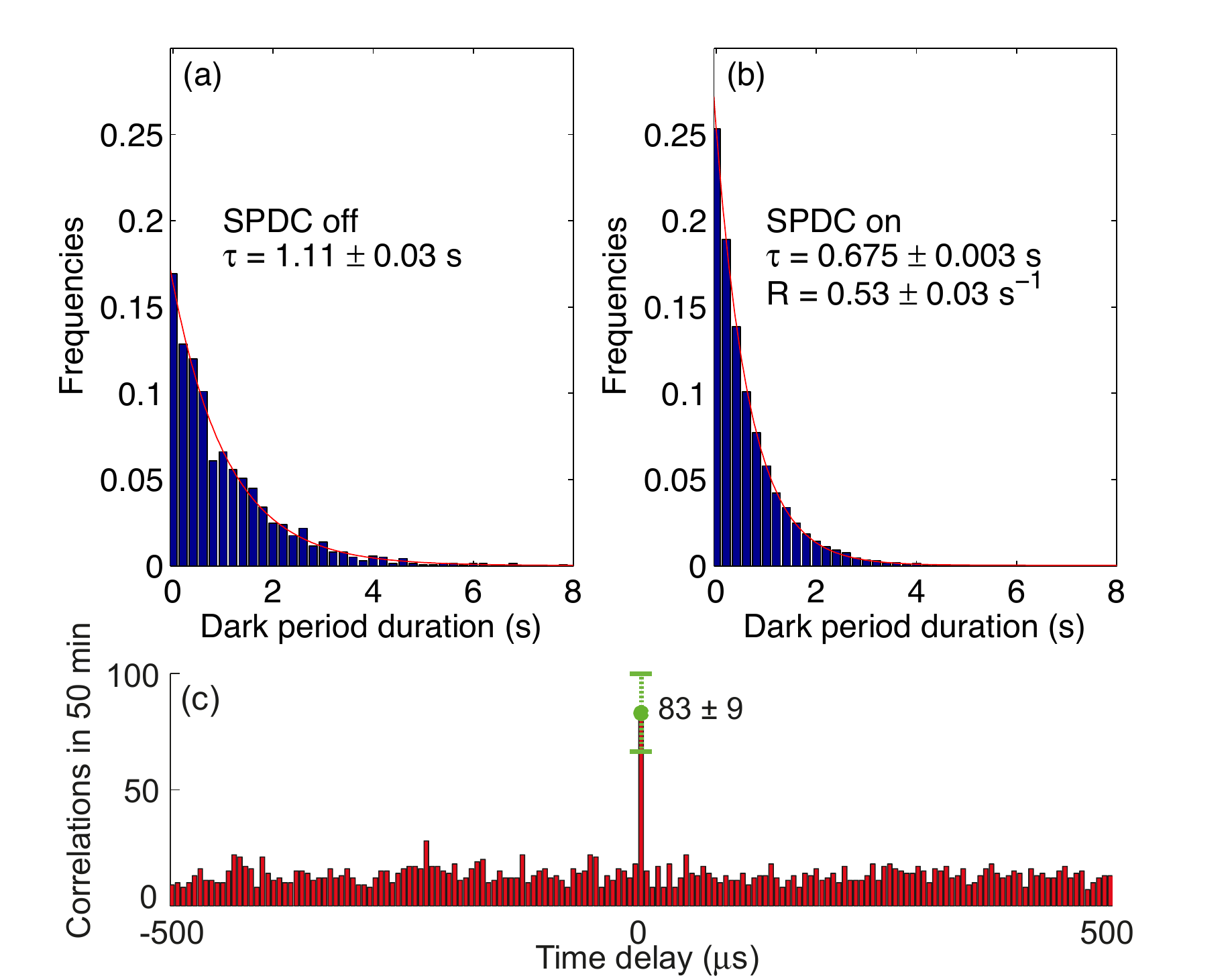}
  \caption{Single photon-single atom interaction data. (a-b) Histograms of dark period durations with the SPDC source off (a) and on (b), showing reduction of the D$_{5/2}$ state population upon exposure to the quantum light source. (c) Second-order correlation function (time delay distribution) between the absorption event (the first photon after a dark-bright quantum jump) and the detection of a herald photon transmitted through the narrowband filter of the SPDC source. \label{fig:DarkPeriodDuration}}
\end{figure}

A simple experiment proves this method. An ion is trapped and illuminated with a 397 nm and 866 nm lasers, while its fluorescence emission is monitored with a photo-multiplier (Figure \ref{fig:IonSourceSetup}). In addition, a weak 850 nm laser resonant with the D$_{3/2}$ to P$_{3/2}$ transition optically pumps the ion (on a 1-s time scale) to the metastable D$_{5/2}$ state, halting the emission of fluorescence. Eventually, the ion will spontaneously decay back to the ground state and restart the fluorescence emission cycle. This is detected in the form of quantum jumps in the fluorescence count trace, see Fig.~\ref{fig:QJdata}a. The corresponding count histogram (Fig.~\ref{fig:QJdata}b) enables defining the optimal count threshold for discrimination of the on and off states, as described in the Appendix. The distribution of dark period durations in the fluorescence count trace, displayed in Fig.~\ref{fig:DarkPeriodDuration}a, exhibits an exponential dependence with a decay constant given by the spontaneous decay rate of the D$_{5/2}$ state, measured to be $\tau_0 = $1.11(3)~s.

When the ion is illumated by a weak 854 nm light source, in our case the SPDC entangled photon source, the ion has two channels to reenter the fluorescing state (see Fig.~\ref{fig:InteractionSchemes}b): either spontaneous decay, or excitation to the P$_{3/2}$ state and decay to the ground state S$_{1/2}$. One should therefore expect a decrease in the lifetime of the metastable D$_{5/2}$ state and, hence, a drop in the average duration of the dark periods in the fluorescence trace. We indeed observe a reduction of the average decay time to $\tau_\mathrm{on}$ = 0.675(3)~s (Fig.~\ref{fig:DarkPeriodDuration}b), and the corresponding absorption rate is derived as $R_\mathrm{spdc} = \tau_\mathrm{on}^{-1} - \tau_\mathrm{off}^{-1} = 0.53(3)$~s$^{-1}$. 

In a subsequent experiment, we prove that the absorption may be attributed to individual SPDC photons. In order to determine the time instant of the absorption event, we extract the first detected fluorescence photon in each dark-bright quantum jump using the technique described in the Appendix. We then calculate the second order correlation function $g^{(2)}(\tau)$ (i.e., the delay time distribution) between this detection and the detection of the heralding SPDC photon (Figure \ref{fig:DarkPeriodDuration}c). A sharp coincidence peak at time delay $\tau = 0$ emerging above the random background proves the strong correlation between the two events and, hence, the heralded absorption of single SPDC photons by the ion. We observe 83(9) coincidences in 50 minutes of acquisition, against a Poisson-distributed background of 13.6(3.7) counts, i.e. at $>22\sigma$ outside the random fluctuations. 

%from which we estimate a probability $p(n\ge83) = 6.7\times 10^{-16}$, that the correlations were obtained due to a random fluctuation of the background.

These experiments prove the possibility of observing and identifying single photon absorption events by a single ion. But to render such interaction useful for quantum networking technologies, a further level of control is necessary: to match the quantum states from the absorbed photon to the atom, the initial Zeeman substate of the atom must be controlled; in addition, the absorption event must be detected without destroying the final state of the ion, which requires detecting the single photon emitted upon the single-photon excitation, as in Figure \ref{fig:InteractionSchemes}(c,d). In the following we show experiments that achieve these two requirements.

\subsection{Polarization control in the absorption event}

To render the absorption process dependent on the polarization of the incoming photon, the atom must be prepared in a given Zeeman substate, or a coherent superposition of two of them, as discussed in Section \ref{subsec:PhotonAtomInteraction}. Experimentally, this is achieved either by optical pumping techniques or by coherent Rabi pulses driving transitions between the relevent levels.

\begin{figure}[t!]
\centering
  \includegraphics[width=\textwidth]{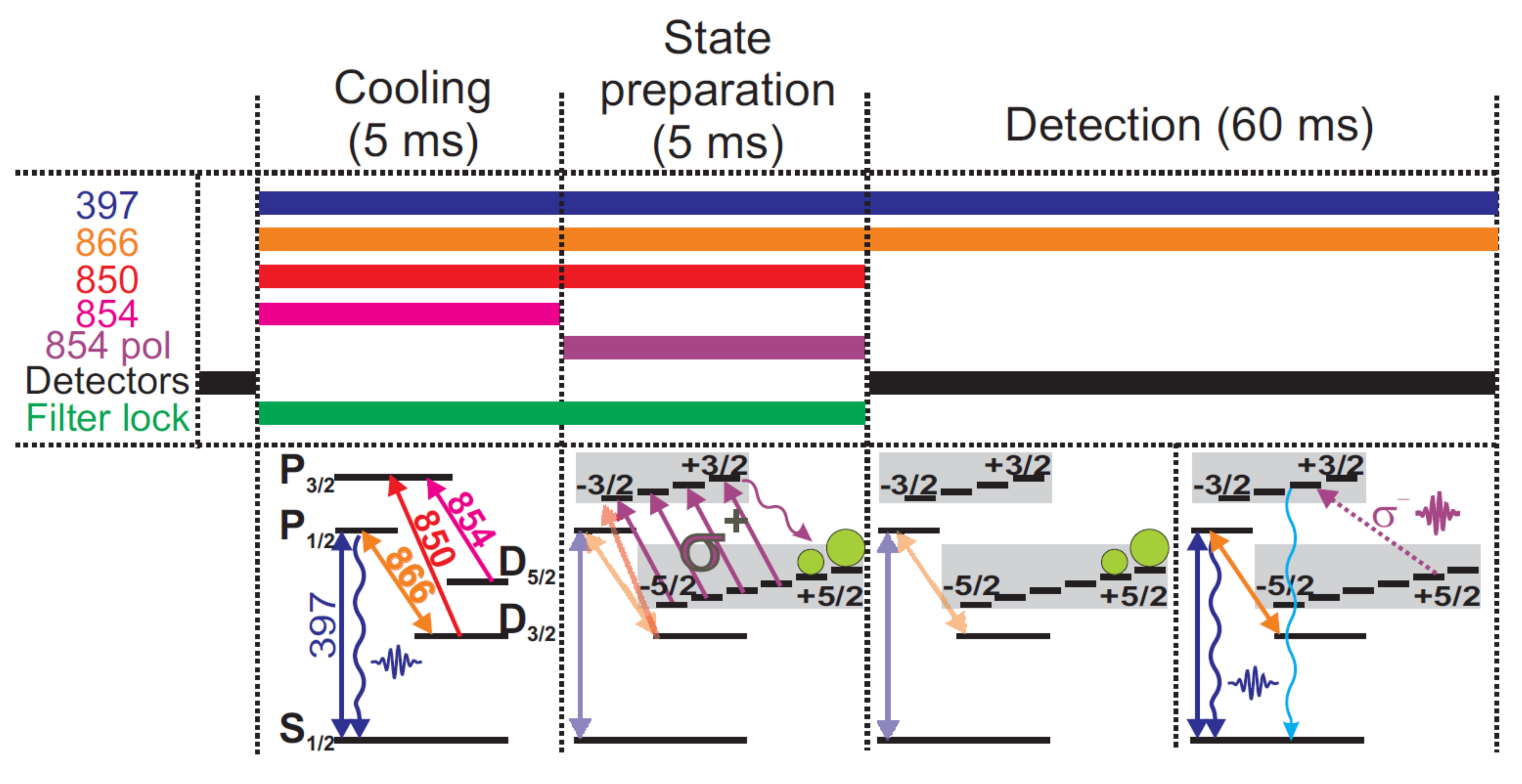}
  \caption{Pulsed sequence used to control the polarization of the absorption process. The ion is laser cooled during 5~ms and then optically pumped to the outer Zeeman substates by means of a circularly polarized 854~nm laser. In the last phase, the pumping laser is turned off and the PMT detectors are activated to monitor fluorescence. Upon a photon absorption event, the ion emits a constant stream of 397 nm fluorescence photons.  \label{fig:PulsedSequence}}
\end{figure}

\begin{figure}[t!]
\centering
  \includegraphics[width=\textwidth]{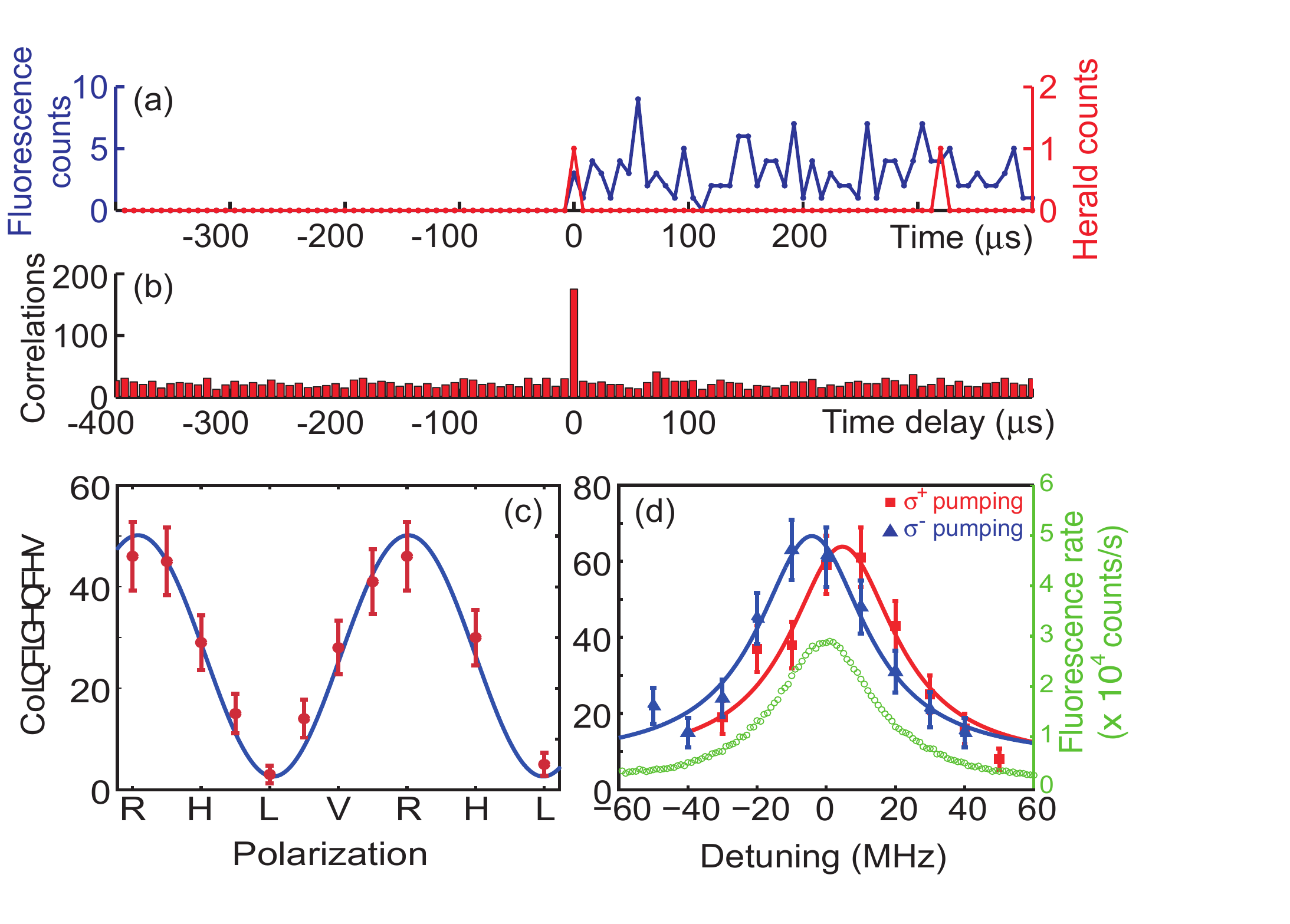}
  \caption{Heralded single-photon absorption with polarization control. (a) Example of a fluorescence onset event (blue trace) coinciding with the detection of a heralding photon from the down-conversion source (red trace). (b) Extracting such events and calculating the correlation function yields a strong coincidence peak at zero time delay, revealing the single-particle character of the interaction. (c) The polarization dependence of the absorption rate is manifested by measuring the coincidence peak for different settings of the input photon polarization (L/R = left-/right-circular, H/V = horizontal/vertical). (d) Frequency dependence of the absorption-herald coincidence rate recorded for pumping to the lower (red points) and upper (blue points) Zeeman substates. A reference absorption spectrum of the D$_{5/2}\leftrightarrow {\textrm P}_{3/2}$ transition (green trace) is measured using fluorescence spectroscopy with a linearly polarized 854~nm laser that propagates along the quantization axis. Solid lines are Lorentzian fits to the data.\label{fig:PulsedCorrelations}}
\end{figure}

To show this, we perform an experiment \cite{Piro2011} similar to the one reported in the last section, using the pulse sequence described in Figure \ref{fig:PulsedSequence}. The ion is first laser cooled by driving it with resonant 866, 850, 854~nm lasers and a red-detuned 397~nm laser. It is then prepared in a mixture of the two outermost Zeeman substates of D$_{5/2}$, by optical pumping with a circularly polarized 854~nm laser, propagating along the quantization axis defined by the applied magnetic field (see Fig.~\ref{fig:IonSourceSetup}). The ion becomes sensitive to either $\sigma^{-}$ polarized photons, if pumped to the upper two states, or $\sigma^{+}$ if pumped to the lower ones. Finally, in the last step, only the 397 and 866~nm lasers are left on and detectors are activated to monitor the ion fluorescence. Strong time correlations are again observed between the first detected fluorescence photon and the SPDC heralding photon (Fig.~\ref{fig:PulsedCorrelations}b). 

Most importantly, the absorption process now displays a strong dependence on the photon polarization. This is observed by recording the height of the coincidence peak for different SPDC photon polarizations, varied from R-circular ($\sigma^-$) to L-circular ($\sigma^+$) (Figure \ref{fig:PulsedCorrelations}c). Full supression of the correlation rate is found when the photon polarization is orthogonal to the one programmed by the optical pumping.

Additionally, a change of the frequency dependence of the absorption with the photon polarization is observed in a single-photon coincidence spectroscopy experiment: the ion is pumped to either the higher or the lower Zeeman substates, and the photon polarization is adjusted to $\sigma^-$ and $\sigma^+$, respectively. The coincidence rate is then recorded for different detunings of the SPDC photons from the atomic transition center frequency (Fig.~\ref{fig:PulsedCorrelations}d). The two single-photon absorption spectra are found to be symmetrically displaced from the center frequency of the transition (determined by fluorescence spectroscopy with a horizontally polarized 854~nm laser), due to the differential Zeeman shift between the involved D$_{5/2}$ and P$_{3/2}$ levels. The spectral line widths of $\sim 45$~MHz fit well with the expectation, resulting from convoluting the atomic transition width with the engineered photon spectral width, each $\sim 22$~MHz.

Note that the single-photon spectroscopy is based on absorption-herald coincidences. The photons sent to the ion have the full SPDC bandwidth of $\sim 200$~GHz, and detuning is only applied to the center frequency of the cavities that filter out the heralds. Due to the strict frequency correlation of the photon pair, however, this selects the events when a photon falls within the resonance bandwidth of the ion. 

These experiments demonstrate heralded single-photon absorption with control on the polarization in the interaction. Nevertheless, the polarization entanglement shared by the SPDC photon pairs is destroyed by splitting them with a polarizing beam splitter. To further develop the potential of our methods for quantum information processing, an experiment was performed where the atom interacts with a photon that is polarization-entangled with its heralding partner, in order to verify that the polarization entanglement is indeed manifested in the absorption process \cite{Huwer2013}. To this end, the photon pairs are now split by means of a non-polarizing beam splitter, such that a coincidence detection between the two arms corresponds to a pair in a maximally entangled polarization state \cite{Haase2009, Piro2009}.

\begin{figure}[t!]
\centering
  \includegraphics[width=\textwidth]{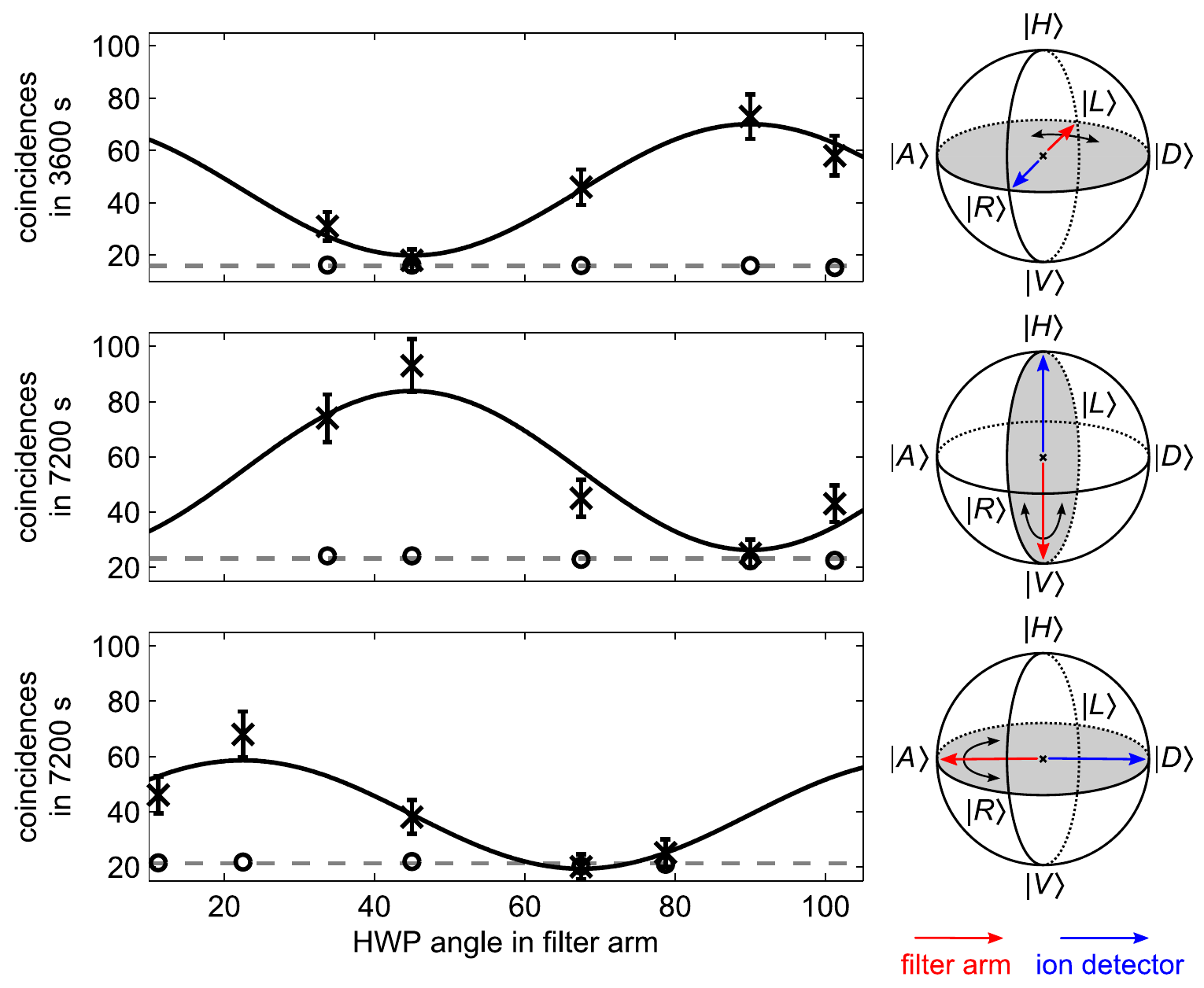}
  \caption{Dependence of absorption-herald coincidences on polarization of the heralding photon for three different polarization bases, R-L (top), H-V (middle), and D-A (bottom). Data points (crosses) are extracted at $\tau = 0$ from the absorption–-herald correlation function $g^{(2)}(\tau)$ on a 10~$\mu$s time grid; the corresponding background (circles), produced by accidental coincidences, is the average over the whole $g^{(2)}(\tau)$ function. Error bars correspond to one standard deviation assuming Poissonian counting statistics. The curves are sinusoidal fits with a fixed period and offset angle. In the right-hand column, we show the Poincar\'e sphere with the setting of the ion (blue) and of the heralding photon detector (red, with black arrows indicating variation by rotating the HWP). From \cite{Huwer2013}.\label{fig:EntanglementPreservingAbsorption}}
\end{figure}

Entanglement is manifested by measuring the polarization correlation, in this case of absorption and herald, in various bases. As for detection of the herald, there is a polarizer in front of the filter cavities, such that a basis rotation is simply performed by an additional quarter- or half-wave plate. Setting the polarization basis of the absorption is more involved; it is attained by adjusting the orientation of the applied magnetic field and the polarization of the 854~nm laser, with respect to the incoming SPDC photon wavevector, during the optical pumping which prepares the ion's state before the absorption. Thereby, the ion is made sensitive to only one photon polarization of the respective linear or circular basis. Correlations are then recorded as a function of the half-wave plate (HWP) angle in the heralding photon arm. As displayed in Fig.~\ref{fig:EntanglementPreservingAbsorption}, they are observed with high visibilities in the three relevant detection basis, left-/right-circular $\left\{\left| R\right\rangle, \left| L\right\rangle\right\}$, horizontal/vertical linear $\left\{\left| H\right\rangle, \left| V\right\rangle\right\}$, and diagonal/anti-diagonal linear $\left\{\left| A\right\rangle, \left| D\right\rangle\right\}$ \cite{Huwer2013}. 

This experiment clearly manifests the polarization entanglement in the single-photon absorption, but the absorbed polarization state is still not preserved in the internal state of the ion, because the latter is destroyed by the continuous 397~nm laser excitation of the fluorescent transition, i.e. by the quantum jump detection technique. The experiment that achieved a full state transfer is described in the following.

\subsection{Photon-to-ion state transfer by heralded absorption}

The protocol which heralds the absorption event while preserving the state of the ion has been described in section \ref{subsec:PhotonAtomInteraction}. It was proposed in \cite{Mueller2014, Sangouard2013} and is related to an earlier scheme for neutral atoms in cavities \cite{Lloyd2001}. The protocol enhances the previous method in two important aspects (see Figure \ref{fig:StateTransferScheme854}). Firstly, absorption happens out of a pure initial state, either a single Zeeman sublevel or a coherent superposition, but not a mixture as before. Secondly, absorption is heralded by the single photon emitted in the Raman transition from the initial D$_{5/2}$ state to the S$_{1/2}$ ground state manifold. 

This has been realized in a recent experiment \cite{Kurz2014}, employing a narrowband laser at 729~nm to coherently excite the S$_{1/2}$ to D$_{5/2}$ transition, plus a radio-frequency field that drives the transition between the two S$_{1/2}$ Zeeman sub-levels. The single absorption event is heralded by detection of the single photon emitted on the P$_{3/2}$ to S$_{1/2}$ transition at 393~nm, without fluorescence excitation. As a result, transfer of the polarization state of photons from a weak laser onto the Zeeman qubit in the S$_{1/2}$ level is demonstrated with over 95\% fidelity, an unprecedented result. Most importantly, the heralding scheme renders the fidelity independent of the success probability.

\section{Conclusions and outlook}\label{sec:ConclusionAndOutlook}

In conclusion, we have reviewed experiments in which a single trapped ion interacts in a controlled manner with single photons. Resonant, heralded single photons are generated for that purpose with an SPDC-based photon pair source. Single interaction events are detected efficiently thanks to the strong amplification provided by the onset of fluorescence upon the absorption of a photon. Absorption events are observed to be strongly time correlated with the time-resolved detection of the partner photon generated simultaneously by the source. The interaction is further controlled by using a laser pulse sequence that optically pumps the ion such that it only absorbs photons of a given polarization. The observed correlation shows almost full supression at the wrong polarization. Extending this technique to polarization-entangled photon pairs and observation in three polarization bases, the photonic entanglement is manifested in the heralded absorption processes. 

The method is then extended into a protocol for mapping the quantum state of a single photon onto the ground-state Zeeman qubit of the ion. For this, the single photon emitted by the ion after the absorption of the incoming photon, i.e. in a Raman scattering process, is used as a heralding signal of the absorption event. The protocol has been demonstrated using laser photons; implementing it with SPDC photons is work in progress. If the incoming photon is provided by an entangled photon pair source, detection of the herald renders the ion entangled with the partner photon. Subsequently, this second photon can be set up to interact with a second distant trapped atom, or even a quantum particle of different nature, resulting in a generic scheme to entangle different massive quantum systems via controlled absorption. Such perspectives become even more attractive in view of recent progress in quantum frequency conversion \cite{Zaske2012}, by means of which single photons may be converted between different wavelength regimes, including telecom bands. Hence, the experiments discussed here are key steps towards making heralded absorption suitable for quantum information processing applications.

Other recent approaches to implement controlled single-photon interaction with a trapped ion involve the direct photonic coupling between two distant ions \cite{Schug2013} and the cavity-enhanced interaction of an ion with photons from a quantum dot \cite{Steiner2014}. Protocols that rely on the interference of photons \textit{emitted} by distant atoms to entangle them have already been succesfully implemented with ions \cite{Moehring2007}, neutral atoms \cite{Lettner2011} and color centers in diamond \cite{Bernien2013}. The entanglement rates of these schemes are limited by the photon collection efficiency into a single optical mode. We believe that the schemes we propose offer a competitive alternative, in that the high generation rates of state-of-the-art narrow-band photon pair sources \cite{Wolfgramm2008, Pomarico2009, Zhou2014} may alleviate the collection efficiency problem.

\begin{acknowledgement}
The experiments described in this chapter were carried out by the members of J. Eschner's group at The Institute of Photonic Sciences (until 2009) and at the Universit\"at des Saarlandes; their contributions are most gratefully acknowledged. We also acknowledge financial support by the MICINN (Consolider 2010 project "QOIT"), the BMBF (projects "QuOReP" and "Q.com-Q", CHIST-ERA project "QScale"), the German Scholars Organization / Alfried Krupp von Bohlen und Halbach-Stiftung, the ESF (COST Action "IOTA"), and CONICYT.
\end{acknowledgement}

\section*{{Appendix: data analysis}}\label{sec:DataAnalysis}

The raw data sets produced by the two types of schemes in Figure \ref{fig:InteractionSchemes} consist of lists of photon detection times from two channels, the source herald and either the ion laser-induced fluorescence (schemes (a) and (b)), or the single emitted photons (schemes (c) and (d)). With the latter, the single-photon interaction is manifested directly by calculating the second-order correlation function between the two detection events. In schemes (a) and (b) where the absorption is signalled by a change in the fluorescent state of the ion (a jump), a previous data processing step has to be undertaken to extract the time of the first (or last) detected fluorescence photon, which is then assigned to be the moment of the quantum jump. Below we provide some detail on how this preprocessing is performed.

The analysis is performed in three steps: 1) determination of the optimal on-off fluorescence threshold; 2) preselection of detection times around each fluorescence jump; 3) detection of the first (last) photon associated to the fluorescence jump. 

Step 1 establishes a criterion to decide whether in a given time interval the ion is in the \textit{on} or \textit{off} fluorescent state. First, the time axis is divided into bins of size $t_\mathrm{b}$, and the fluorescence trace, i.e. the number $n_\mathrm{f} (i)$ of detections in bin $i$, is computed. From the trace, a histogram of fluorescent counts is extracted. This histogram statistically corresponds to the sum of two independent Poisson distributions $P_\mathrm{on/off}(\bar{n}_\mathrm{on/off},n)$, with mean values $\bar{n}_\mathrm{on/off}$ given by the average detection rate for the on and off states in the specified time bin size. The optimal threshold used to discriminate the two states is then given by $n_\mathrm{th} = \min_{n} \left\{ P_\mathrm{on}(\bar{n}_\mathrm{off},n) + P_\mathrm{off}(\bar{n}_\mathrm{on},n) \right\} $, with $\bar{n}_\mathrm{off} > n > \bar{n}_\mathrm{on}$.

Step 2 is performed by running a moving average over the fluorescence trace, $\bar{n}_\mathrm{f} (i)$ of window size $N$, given by the desired number of detection events to be extracted, storing the last $N$ detection times in a buffer. Every time the condition $\bar{n}_\mathrm{f} (i) < n_\mathrm{th} < \bar{n}_\mathrm{f} (i\pm1)$ is met (using $+$ ($-$) sign for detecting the change to the \textit{on} (\textit{off}) fluorescing state) the last $N$ stored detection times are extracted.

Step 3 is performed directly on the detection times $t_i$ extracted in Step 2. They are scanned for the condition $t_i > \tau_\mathrm{th} > t_{i\pm1} $, where the sign is chosen to detect the first (+) or last (-) fluorescence photon. The optimal delay threshold $\tau_\mathrm{th}$ to discriminate the \textit{on} and \textit{off} state is chosen by maximizing the probability of succesfully detecting an atom transition, namely, the probability of \textit{not} detecting a photon in the \textit{off} state after a time $\tau_\mathrm{th}$ and detecting one in the \textit{on} state before a time $\tau_\mathrm{th}$. Assuming a rate of fluorescence photon detection $r_\mathrm{on}$ and a dark count rate of $r_\mathrm{off}$ in the \textit{off} state, this probability is given by $p_\mathrm{det} =\exp\left(-r_\mathrm{off} \tau_\mathrm{th}\right)\left[1-\exp(-r_\mathrm{on}\tau_\mathrm{th})\right]$, which takes a maximum value at 
$\tau_\mathrm{th} = r_\mathrm{on}^{-1}\log \left(1+r_{\mathrm{on}}/r_{\mathrm{off}}\right)$.

\addcontentsline{toc}{section}{Appendix}
%
%

%\section{References}
%We strongly recommend using \keyword{BibTeX} to handle references, as done in this template.  Use the \verb|cite| %command to generate in-text \keyword{citations}, e.g. %\cite{BraunsteinBOOK2010,kuzmich2003atomic,PredojevicCLEQEC2009,WolfgrammPRL2011}

%\bibliographystyle{spphysMWM}
\bibliographystyle{SpringerPhysMWM} %Modification of nature.bst by MWM
\bibliography{Literature}

\printindex
\end{document}